\documentclass[10pt]{article}

\usepackage{amssymb}
\usepackage{graphicx}
\usepackage[numbers,compress]{natbib}
\begin{document}
	
	\title{Null energy condition of dynamical black holes  in spatially flat FLRW space-times}	

	\author{Ion I. Cot\u aescu\thanks{Corresponding author E-mail:~~i.cotaescu@e-uvt.ro},\\
		{\it West University of Timi\c soara,} \\{\it V. Parvan Ave. 4,
			RO-300223 Timi\c soara}}

\maketitle

\begin{abstract}
A new identity reported here for the first time is used for studying how the null energy condition can be applied to the models of dynamical black holes proposed recently   [I. I. Cot\u aescu, Eur. Phys. J. C  (2022) 82:86].   It is shown that this condition selects the models with physical meaning whose horizons evolve in accordance with the general black hole theory in frames with null coordinates [S. A. Hayward, Phys. Rev. D {\bf 49} (1994) 6467].  

Pacs: 04.70.Bw
\end{abstract}

Keywords:  dynamical black holes; asymptotic FLRW spece-time; null energy condition; dynamical horizons; black hole evaporation.

\section{Introduction}

For building models of static black holes (see for instance Ref. \cite{BH})  populating  Friedmann- Lema\^ itre-Robertson-Walker (FLRW) universes  external gravitational sources are needed as, for example, the cosmological constant used in the case of the Schwarzschild-de Sitter black hole (having the Kottler metric \cite{Kot}). Such inconveniences encouraged many authors to look for dynamical models \cite{TB1} describing coherently the evolution of the black holes in their  asymptotic FLRW backgrounds, without supplemental hypotheses.  The first model of dynamical particle behaving as a black hole was proposed by McVittie long time ago \cite{MV} and then studied and generalized by many authors \cite{MV1,MV2,MV3,MV4}.  The general feature of these models is that their gravitational source is the fluid pressure which is singular on the Schwarzschild sphere while the fluid density remains just that of the asymptotic FLRW space-time.

Alternatively,  we proposed recently  a new type of dynamical particles which are exact solutions of Einstein's equations with perfect fluid but this time preserving the pressure of the perfect fluid of the asymptotic FLRW space-times  \cite{Cot,Cot1}. These models are determined by a time-dependent black hole mass and a real valued parameter $\kappa$ such that we say that these are $\kappa$-models \cite{Cot1}. The solutions we studied previously \cite{Cot} are  particular such models with $\kappa=0$.   We have shown that these models describe systems constituted by a Schwarzschild-type black holes surrounded by a cloud of dust, hosted by the perfect fluid which determines the FLRW asymptotic behaviour of the background.  When the  background  is expanding then the black hole evaporates dissipating its mass into dust but without affecting the FLRW fluid \cite{Cot2}.  In these models the Hubble function of the FLRW space-time is proportional through $\kappa$ to the total mass of the dynamical system black hole dust such that we do not need to consider a cosmological constant in the case of the evaporating black holes in de Sitter expanding universe \cite{Cot2}. 

The $\kappa$-models describe dynamical black holes only  in the physical space domain bordered by the black hole and cosmological dynamical horizons. In the examples studied so far \cite{Cot,Cot1,Cot2} we have shown that, in expanding universes, these horizons behave naturally, the black hole one collapsing to zero while the cosmological horizon is increasing tending asymptotically to the apparent horizon of the FLRW asymptotic space-time.  On the other hand, we know that the general theory of black holes in local frames with null coordinates \cite{H1,H2} predicts a black hole dynamics which depends on the null energy condition. We tested this condition only in the particular case studied in Ref, \cite{Cot2} where we calculated the null energy showing that this is non-negative, but at that time we did not have a general rule for evaluating the null energy of the $\kappa$-models. Fortunately, in the meanwhile we deduced an important identity allowing us to derive the null energy of all our $\kappa$-models. The principal goal of this letter is to report this identity discussing its role in selecting the models with physical meaning. After presenting briefly our models of dynamical black holes in the next section, we focus in Sec. 3 on the general form of the null energy condition deriving the rules selecting the models satisfying this condition.  Finally, we present two short remarks in the last section.

We use the Planck units with $\hbar=c=G=1$.

\section{Models of dynamical black holes}

Let us start revisiting the principal features of our $\kappa$-models describing dynamical non-rotating black holes  in space-times with spherical symmetry and spatially flat sections \cite{Cot1,Cot2}. For these models we use physical frames, $\{t, {\bf x}\}$,  with  Painlev\' e-Gullstrand  coordinates  \cite{Pan,Gul},  $x^{\mu}$ ($\alpha,\mu,\nu,...=0,1,2,3$) formed by the {\em cosmic time}, $x^0=t$, and physical Cartesian space coordinates,  ${\bf x}=(x^1,x^2,x^3)$, associated to the spherical ones $(r,\theta,\phi)$. The  line elements  in physical frames have the general form  
\begin{eqnarray}
	ds^2&=& g_{\mu\nu}(x)dx^{\mu}dx^{\nu}=dt^2 -\left[dr-h (t,r)dt\right]^2-r^2 d\Omega^2 \nonumber\\
	&=&\left[1-h (t,r)^2\right]dt^2+2 h(t,r)dr dt -dr^2-r^2 d\Omega^2\,,\label{fam}
\end{eqnarray} 
depending on the smooth functions $h$ and $d\Omega^2=d\theta^2+\sin^2\theta\, d\phi^2$. 

In what follows we focus on the $\kappa$-models \cite{Cot1} whose $h$-functions in proper co-moving frames where the four-velocity has the components 
\begin{eqnarray}
	U_{\mu}&=&\left(\frac{1}{\sqrt{g^{00}}},0,0,0\right)=(1,0,0,0)\,,\label{U1}\\ 
	U^{\mu}&=&g^{\mu 0}U_0=\frac{g^{\mu 0}}{\sqrt{g^{00}}} ~~\Rightarrow~~U^{\mu}=(1,h,0,0)\,,\label{U2}
\end{eqnarray}
have the form
\begin{equation}\label{hdef}
	h_{\kappa}(t,r)=-\frac{1}{3}\frac{\dot{M}(t)}{M(t)} r+\epsilon\sqrt{\frac{2 M(t)}{r}+\kappa^2 M(t)^2 r^2}\,,
\end{equation}
depending on the non-negative dynamical mass, $M(t)>0$, its time derivative, $\dot M(t)=\partial_t M(t)$,  and the  real valued constant $\kappa=\epsilon|\kappa|$  ($\epsilon={\rm sign}(\kappa)$) playing the role of free parameter. The space-times of these models,  ${\frak M}(M,\kappa)$,  have isotropic Einstein tensors which satisfy Einstein's equations, 
\begin{equation}
	G^{\mu}_{\,\nu}(M,\kappa)=8\pi [ (\rho_{\kappa}+p_{\kappa})U^{\mu}U_{\nu}-p_{\kappa}\delta^{\mu}_{\nu}]\,,
\end{equation}
with a perfect fluid of density $\rho_{\kappa}$, pressure $p_{\kappa}$ and four-velocities of components (\ref{U1}) and (\ref{U2}),   in the proper co-moving physical frame. Hereby we obtain the diagonal components \cite{Cot1} 
\begin{eqnarray}
	G^0_0(M,\kappa)&=&3\kappa^2 M(t)^2 +\frac{1}{3} \frac{\dot M(t)^2}{M(t)^2}-2\dot{M}(t)K(t,r)\nonumber\\
	&=&8\pi \rho_{\kappa}\,,\label{E1}\\
	G(M,\kappa)&\equiv&G^r_r(M,\kappa)=G^{\theta}_{\theta}(M,\kappa)=G^{\phi}_{\phi}(M,\kappa)\nonumber\\
	&=&3\kappa^2 M(t)^2+ \frac{\dot M(t)^2}{M(t)^2}-\frac{2}{3} \frac{\ddot M(t)}{M(t)}=-8\pi p_{\kappa}	\,,\label{E2}
\end{eqnarray}
where 
\begin{equation}\label{K}
	K(t,r)=\epsilon\frac{1+{M(t)}\kappa^2r^{3} }{\sqrt{M(t)(M(t)\kappa^2r^3+2) r^3}}\,.	
\end{equation}
The only non-vanishing off-diagonal component of the Einstein tensor is  
\begin{eqnarray}
	G^r_0(M,\kappa)&=&8\pi(\rho_{\kappa}+p_{\kappa})U^{r}U_{0}\nonumber\\
	&=&\frac{g^{0r}}{g^{00}}\left(G^0_0(M,\kappa)-G(M,\kappa)\right)\,.\label{cond}
\end{eqnarray}

The asymptotic space-time of  ${\frak M}(M,\kappa)$, for $r\to\infty$,  is a FLRW manifold  with a scale factor $a(t)$, denoted by ${\frak M}(a)$, whose Hubble function is defined by the asymptotic condition \cite{Cot1}
\begin{equation}\label{hubb}
	\frac{\dot a(t)}{a(t)}=\lim_{r\to\infty}\frac{h_{\kappa}(t,r)}{r}=\kappa M(t) -\frac{1}{3}\frac{\dot{M}(t)}{M(t)}\,.
\end{equation}
The metric tensors of  ${\frak M}(a)$, give the asymptotic line elements in physical frames of the form,
\begin{equation}\label{s2}
	ds^2=\left(1-\frac{\dot a^2}{a^2}\, r^2\right)dt^2+2\frac{\dot a}{a}\, r\, dr\, dt -dr^2-r^2d\Omega^2\,,
\end{equation}
emphasizing the apparent horizon on the sphere of radius  
\begin{equation}\label{ra}
	r_a(t)=\left| \frac{a(t)}{\dot a(t)}\right| \,,
\end{equation}
we call the asymptotic horizon. 
Integrating Eq. (\ref{hubb})  with the initial conditions $a(t_0) =1$ and $M(t_0)=M_0$ we obtain the expression of the scale factor  
\begin{equation}\label{at1}
	a(t)=\left(  \frac{M_0}{M(t)}\right)^{\frac{1}{3}} \exp\left(\kappa\int_{t_0}^t M(t')dt'\right)
\end{equation}
of the asymptotic FLRW space-time, in terms of mass function and $\kappa$. Once we have the scale factor $a(t)$ we may derive the asymptotic density and pressure, $\rho_a$ and respectively $p_a$, resulted from the Friedmann equations, 
\begin{eqnarray}
	\rho_a(t)&=&\frac{3}{8\pi}\left(\frac{\dot a}{a}\right)^2  \nonumber\\
	&=&\frac{1}{8\pi}\left( 3\kappa^2 M(t)^2 +\frac{1}{3} \frac{\dot M(t)^2}{M(t)^2}-2\kappa \dot M(t)\right)\,,\\
	p_a(t)&=&- \frac{1}{8\pi}\left[ 3\left(\frac{\dot a}{a}\right)^2 +2\frac{d}{dt}\left(\frac{\dot a}{a}\right)   \right]  \nonumber\\
	&=&-\frac{1}{8\pi}\left( 3\kappa^2 M(t)^2+ \frac{\dot M(t)^2}{M(t)^2}-\frac{2}{3} \frac{\ddot M(t)}{M(t)} \right)\,,
\end{eqnarray}
which depend only on the Hubble function (\ref{hubb}). Hereby it results that the total density $\rho_{\kappa}=\rho_a + \delta\rho$  of the perfect fluid of the space-time ${\frak M}(M,\kappa)$ gets the new point-dependent  term \cite{Cot1}
\begin{equation}
	\delta\rho(t,r)=\frac{1}{4\pi} \dot M(t)\left[  \kappa  - K(t,r)   \right] \,, \label{dust}	
\end{equation}
while its pressure remains unchanged, $p_{\kappa}=p_a$. This means that $\delta\rho$ is the density of an amount of {\em dust} which does not modify te pressure $p_a$ of the FLRW fluid.  

The special case of $\kappa=0$ must be considered separately starting with functions of the form 
\begin{equation}\label{fun}
	\hat h_{\epsilon}(t,r)=\frac{\dot a(t)}{a(t)} r +\epsilon \sqrt{\frac{2\hat M(t)}{r}}\,, \quad \hat M(t)=\frac{M_0}{a(t)^3}\,,
\end{equation} 
where we keep the parameter $\epsilon=\pm 1$ and the above initial condition $M_0=\hat M(t_0)$.  These functions give the metric tensors of the space-times ${\frak M}(\hat M)$  having isotropic  Einstein tensors defining the density $\hat\rho_{\epsilon}=\rho_a+\delta\hat\rho$ and pressure $\hat p_{\epsilon}=p_a$ of the perfect fluid where
\begin{equation}\label{fuk}
	\delta\hat\rho(t,r)=  3\epsilon\frac{\dot a(t)}{a(t)}\sqrt{\frac{2 \hat M(t)}{r^3}}\,,	
\end{equation} 
represents the dust density. In Ref. \cite{Cot} we have shown that the models with $\epsilon=1$ describe dynamical black holes in expanding space-times.

In the space-times ${\frak M}(M,\kappa)$ or ${\frak M}(\hat M)$, the coordinate $t$ of the physical frames  is the cosmic time only in the physical domains where  $g_{00}(t,r)>0$. This means that the physical space domain is delimited by two dynamical spherical horizons, the {\em black hole} and {\em cosmological} ones whose radii, $r_b(t)$ and respectively $r_c(t)$,  have to be derived solving the equation $h_{\kappa}(t,r)=\epsilon$ (or $\hat h_{\epsilon}(t,r)=\epsilon$ if $\kappa=0$ ). In many cases  these equations  have the desired positive solutions  only after a critical instant $t_{cr}$ when the horizons arise simultaneously on same sphere of radius $r_b(t_{cr})=r_c(t_{cr})$.  We have seen that in expanding manifolds we studied so far \cite{Cot,Cot1,Cot2} the black hole horizon is decreasing collapsing to zero while the cosmological horizon increases approaching asymptotically to the asymptotic horizon.  For the collapsing space-times we do not have yet examples concerning the horizons dynamics but we know that the black hole models with physical meaning must satisfy the null energy condition \cite{H2}.  

\section{Null energy condition}

Giving the stress-energy tensor $T_{\mu\nu}$,  the null energy is defined as  ${\cal E}=T_{\mu\nu}n^{\mu}n^{\nu}$ where $n$  is a null vector, $n^{\mu}n_{\mu}=0$. In the case of our $\kappa$-models, after a few manipulation, we are surprised to find that the null energy  satisfies the remarkable identity
\begin{equation}\label{null}
	{\cal E}_{\kappa}(t,r)=\rho_{\kappa}(t,r)+p_{\kappa}(t)=-\frac{1}{4\pi\,r}\partial_t h_{\kappa}(t,r)\,,
\end{equation}
which is the principal new result we report here. For $r\to \infty$ we use Eq. (\ref{hubb}) for deriving  the well-known asymptotic null energy
\begin{equation}
	{\cal E}_a(t)=\lim_{r\to \infty} {\cal E}_{\kappa}(t,r)=\rho_{a}(t)+p_{a}(t)=-\frac{1}{4\pi}\frac{d}{dt}\frac{\dot a(t)}{a(t)}\,,
\end{equation}
of the asymptotic FLRW background. Note that from Eq. (\ref{ra}) we may deduce
\begin{eqnarray}
	-\frac{d}{dt}\frac{\dot a(t)}{a(t)}=\left\{ 
	\begin{array}{lll}
		-\frac{d}{dt}\left(\frac{1}{r_a(t)} \right)=\frac{\dot r_a(t)}{r_a(t)^2}\ge 0&{\rm for}&\epsilon=1\,,\\
		\frac{d}{dt}\left(\frac{1}{r_a(t)} \right)=-\frac{\dot r_a(t)}{r_a(t)^2}\ge 0&{\rm for}&\epsilon=-1\,,	
	\end{array}
	\right.
\end{eqnarray} 
which means that the asymptotic FLRW space-times ${\frak M}_a$ satisfy the null energy condition in both expanding ($\dot r_a(t)>0$) or collapsing ($\dot r_a(t)<0$) cases if we assume that the evolution is monotonous such that the functions $r_a(t)$ do not have zeros during the evolution. These results complete our framework giving us the opportunity of selecting models describing dynamical black holes whose dynamical black hole apparent horizons are trapping surfaces  evolve according to the general rules (See Theorem 3 of Ref. \cite{H2}). 

For selecting the desired models we observe that the Hubble function of the asymptotic FLRW space-time must also evolve monotonously in time, without zeros in the physical time domain, $t>t_{cr}$, which  might produce singularities of the functions $r_a(t)$.   We may prevent  these zeros to appear imposing different conditions  for expanding or collapsing  space-times as 
\begin{eqnarray}
	{\rm expanding:} &~~~~&\frac{\dot a(t)}{a(t)}>0 ~~\Rightarrow ~~ 	\frac{\dot{M}(t)}{M(t)}<0\,,\quad \epsilon=1\,,\label{expand}\\
	{\rm collapsing:} &~~~~&	\frac{\dot a(t)}{a(t)}<0 ~~\Rightarrow ~~ 	\frac{\dot{M}(t)}{M(t)}>0\,,\quad \epsilon=-1\,.\label{collaps}
\end{eqnarray} 
In other respects,  from Eq. (\ref{K}) we deduce that $|K(t,r)|\ge|k|$ the dust mass  (\ref{dust}) is non-negative,  $\delta\rho(t,r)\ge 0$, and we have 
\begin{equation}
	{\cal E}_{\kappa}(t,r)=	{\cal E}_{a}(t)+\delta\rho(t,r)\ge 	{\cal E}_{a}(t)\,.
\end{equation} 
This means that the null energy condition ${\cal E}_{\kappa}(t,r)\ge 0$ is 
accomplished if  the background evolves monotonously having ${\cal E}_{a}(t)\ge 0$. For deriving the equivalent conditions for the mass functions $M(t)$ we consider the identity   
\begin{equation}
	\frac{d}{dt}\frac{\dot a(t)}{a(t)}=\frac{d}{dt}\left(\kappa M(t) -\frac{1}{3}\frac{\dot{M}(t)}{M(t)} \right)\le 0\,,
\end{equation}
resulted from Eq. (\ref{hubb}), observing that the both terms must have the sane signs for avoiding the zeros of this expression which is supposed to evolve monotonously.   Combining then this result with the restrictions (\ref{expand}) and (\ref{collaps}) we obtain two sets of equivalent selection rules in the case of expanding space-times with $\kappa>0$ ($\epsilon=1$),
\begin{equation}\label{res}
	\frac{\dot{M}(t)}{M(t)}\le 0\,,~~\frac{d}{dt}\frac{\dot M(t)}{M(t)}\ge 0 ~~ \Leftrightarrow~~\frac{\dot a(t)}{a(t)}\ge 0\,,~~ 		\frac{d}{dt}\frac{\dot a(t)}{a(t)}\le 0 \,, 
\end{equation}
as well as for collapsing space-times with $\kappa<0$ ($\epsilon=-1$),
\begin{equation}\label{res1}
	\frac{\dot{M}(t)}{M(t)}\ge 0\,,~~\frac{d}{dt}\frac{\dot M(t)}{M(t)}\ge 0 ~~ \Leftrightarrow~~\frac{\dot a(t)}{a(t)}\le 0\,,~~ 		\frac{d}{dt}\frac{\dot a(t)}{a(t)}\le 0 \,. 
\end{equation}
These restrictions guarantee that the manifold ${\frak M}(M,\kappa)$ and its asymptotic space-time ${\frak M}(a)$ satisfy simultaneously the null energy condition. For the models with $\kappa=0$, we discussed in Ref. \cite{Cot},  these conditions are fulfilled  because of the special forms of the functions (\ref{fun}) and (\ref{fuk}). For $\kappa\not=0$ the selection rules (\ref{res}) and (\ref{res1}) are useful tools in selecting the models as there are many $\kappa$-models which do not comply with these requirements. For example, the model with the mass function $M(t)\propto e^{-t^2}$ and $\kappa>0$ violates Eqs. (\ref{fun}) and must be eliminated.

Now we can verify that all the expanding $\kappa$-models we studied previously satisfy the null energy condition. In Ref. \cite{Cot1} we considered two type of  examples, either models with mass functions $M(t)\propto t^{-s}$ with $s>0$ or models having asymptotic scale factors  $a(t)\propto t^p$ with $p>0$.  Obviously,  these models  satisfy the restrictions (\ref{res}) for $t>t _{cr}>0$,  complying with the null energy condition as well as the model of  Ref. \cite{Cot2} for which we calculated this condition explicitly. This explains why the horizons of all these models  behave naturally as predicted by the general theory, the black hole (or future inner) horizon being non-increasing while the cosmological (or future outer) horizon is non-decreasing.

\section{Concluding remarks}

The general conclusion is that our $\kappa$-models with $\kappa>0$ which satisfy the null energy condition  have dynamical masses decreasing monotonously  in time in expanding backgrounds.   These systems may be interpreted as evaporating black holes whose masses are dissipating into  dust adsorbed by the perfect fluids of the asymptotic FLRW space-times \cite{Cot1,Cot2}. 

The problem which remains open is of the $\kappa$-models in collapsing space-times which may have $\kappa<0$ and increasing mass functions. We hope that the null energy condition we derived here will help us to select such models with physical meaning and horizon dynamics in accordance with the general black hole theory  \cite{H2}.


\begin{thebibliography}{99}
	
	\bibitem{BH}
	V. P. Frolov and A. Zelnikov, {\em Introduction to Black Hole Physics} (Oxford Univ. Press. Inc., New York 2011).	
	
	\bibitem{Kot}
	F. Kottler,  Ann. Phys. (Berlin) {\bf 361} (1918) 401.	
	
	\bibitem{TB1}
	C. Gao, X. Chen, Y.-G. Shen, and V. Faraoni, Phys. Rev.
	D 84, 104047 (2011).
	
	\bibitem{MV}
	G. C. McVittie, {\em  MNRAS} {\bf 93} (1933) 325.
	
	\bibitem{MV1}
	B. C. Nolan, {\em Phys. Rev. D} {\bf 58} (1998) 064006.
	
	\bibitem{MV2}
	M. Carrera and D. Giulini, {\em Phys. Rev. D} {\bf 81} (2010) 043521.
	
	\bibitem{MV3}
	M. Carrera and D. Giulini, {\em Rev. Modern Phys.} 82 (2010) 169.
	
	\bibitem{MV4}
	R. Nandra, A. N. Lasenby and M. P. Hobson, MNRAS {\bf 422} (2012) 2931.
	
	\bibitem{Cot}
	I. I. Cot\u aescu, Eur. Phys. J. C {\bf 82} (2022) 86. 	
	
	\bibitem{Cot1}
	I. I. Cot\u aescu, Eur. Phys. J. C {\bf 84} (2024) 819.
	
	\bibitem{Cot2}
	I. I. Cot\u aescu, Eur. Phys. J. C {\bf 85} (2025) 318.
	
	
	\bibitem{Pan}
	P. Painleve, C. R. Acad. Sci. (Paris) {\bf 173} (1921) 677.
	
	\bibitem{Gul}
	A. Gullstrand, Arkiv. Mat. Astron. Fys. {\bf 16} (1922) 1.
	
	\bibitem{H1}
	S. W. Hawking and G. F. R. Ellis, {\em The Large Scale Structure of Space Time}  (Cambridge University Press, Cambridge, England, 1973).
	
	\bibitem{H2}
	S. A. Hayward, Phys. Rev. D {\bf 49} (1994) 6467.
	
	
	
\end{thebibliography}
\end{document}